\documentstyle[12pt]{article}
\newcommand\be{\begin{equation}}
\newcommand\ee{\end{equation}}
\newcommand\bea{\begin{eqnarray}}
\newcommand\eea{\end{eqnarray}}
\newcommand\ket[1]{|#1\rangle}

\newcommand\braket[2]{\langle #1|#2\rangle}
\newcommand{\fatalpha}{{\bf \alpha \kern -0.44em \alpha}}
\newcommand{\fatsigma}{{\bf \sigma \kern -0.54em \sigma}}
\newcommand{\tpchi}{{\bf \chi \kern -0.35em \chi}}
\newcommand{\llambda}{{\bf \lambda \kern -0.45em \lambda}}

\bibliography{plain}
\title{\bf Quantum search in structured database using local adiabatic evolution and spectral methods}\vspace{20mm}
\author{ R. Sufiani$^{a,b}$
  \thanks{E-mail:sofiani@tabrizu.ac.ir},
  N. Bahari$^{a}$
 \\ $^a${\small Department of Theoretical Physics and Astrophysics,
University of Tabriz, Tabriz 51664, Iran.} \\ $^b${\small Institute
for Studies in Theoretical Physics and Mathematics, Tehran
19395-1795, Iran.},\\ University of Tabriz, Tabriz 51664, Iran.}
\pagebreak

\vspace{20mm}
\begin{document}
\maketitle \vspace{15mm}
\newpage
\begin{abstract}
Since Grover's seminal work which provides a way to speed up
combinatorial search,  quantum search has been studied in great
detail. We propose a new method for designing quantum search
algorithms for finding a "marked" element in the state space of a
graph. The algorithm is based on a local adiabatic evolution  of the
Hamiltonian associated with the graph. The main new idea is to apply
some techniques such as Krylov subspace projection methods, Lanczos
algorithm and spectral distribution methods. Indeed, using these
techniques together with the second-order perturbation theory, we
give a systematic method for calculating the approximate search time
at which the marked state can be reached. That is, for any
undirected regular connected graph which is considered as the state
space of the database, the introduced algorithm provides a
systematic and programmable way for evaluation of
the search time, in terms of the corresponding graph polynomials.\\


{\bf PACs Index: 03.65.Ud }
\end{abstract}

\vspace{70mm}
\newpage
\section{Introduction}
Grover's quantum search algorithm \cite{Grov} is one of the main
applications of quantum computation. This algorithm is sometimes
described as a way for searching a marked item in an unstructured
database of $N$ items in time $O(\sqrt{N})$. This gives a quadratic
speedup over the exhaustive search for a variety of search problems
\cite{amb1}. But the algorithm as originally proposed is not
designed to search a physical database. In \cite{amb2}, Aaronson and
Ambainis present a model of query complexity on graphs, where they
showed that a database of $N$ items laid out in $d$ spatial
dimensions can be searched in time of order $\sqrt{N}$. In
\cite{childs1}, Childs and Goldstone have considered an alternative
quantum search algorithm based on a continuous time quantum walk on
a graph. Quantum walks provide a natural framework for the spatial
search problem because the graph can be used to model the locality
of the database. In fact, for the case of the complete graph
(unsorted database), the resulting algorithm is simply the
continuous time search algorithm of Farhi and Gutmann \cite{Farhi1}.
On the hypercube, their results showed that the algorithm also
provides quadratic speedup \cite{Farhi2,Farhi3,Farhi4}. In
\cite{Farhi2}, Farhi et al. have used the time-dependant Hamiltonian
approach for Grover's problem, where they have considered the
constant-rate adiabatic approach (the requirement of adiabaticity is
expressed globally) and obtained a complexity of order $N$, the
number of items in database. Then, Roland and Cerf \cite{Roland}
have been considered the same problem with the same approach with
this deference that they considered the adiabatic evolution locally,
i.e., they adjusted the evolution rate of the Hamiltonian so as to
keep the evolution adiabatic on each infinitesimal time interval and
by this adjusting the total running time of order $\sqrt{N}$ has
obtained. Recently, another adiabatic version of the quantum search
problem has considered by Rezakhani et al. in \cite{reza}, where
they have employed continuous time, global adiabatic evolution in
order to calculate a quantity called ``adiabatic error", which
quantifies the distance between the instantaneous ground state and
the actual marked state.

In this paper, we follow the approach of the paper \cite{Roland} and
employ some techniques such as Krylov subspace projection methods,
Lanczos algorithm, spectral distribution methods and the
second-order perturbation theory in order to give a systematic
method for calculating the total search time approximately. In fact,
by reducing the Hilbert space of the corresponding Hamiltonian to
the smaller one called Krylov subspace and using the spectral
methods, we calculate the minimum energy gap between two lowest
eigenvalues of the Hamiltonian and consequently the total search
time, in terms of the polynomials associated with the graph.
 The organization of the paper is as follows: In section
2, we recall some preliminary facts about Krylov subspace projection
methods and spectral distribution method needed for the approach of
the paper. In section 3, by using the local adiabatic evolution
approach, quantum search on a graph is investigated, where an
analytical but approximate formula for the quantum search time (the
main result of the paper) is obtained in terms of the graph
polynomials. Section 4 is devoted to some examples of graphs, in
order to clarify the introduced method in details. The last section
contains a brief conclusion.
\section{Krylov subspace and spectral methods}
In this section we give a brief review of  some of the main features
of Krylov subspace projection methods and spectral distribution
method (the reader is referred to Refs. \cite{bpa}-\cite{lipkin} for
more details).
\subsection{Krylov subspace projection methods}
Krylov subspace projection methods (KSPM) are probably the most
important class of projection methods for linear systems and for
eigenvalue problems. In  KSPM, approximations to the desired
eigenpairs of an $n\times n$ matrix $A$ are extracted from a
$d$-dimensional Krylov subspace
\begin{equation}
 K_d(\ket{\phi_0},A) = span\{\ket{\phi_0},A\ket{\phi_0}, \cdots,A^{d-1}\ket{\phi_0}\},
 \end{equation}
  where $\ket{\phi_0}$
is often a randomly chosen starting vector called reference state
and $d \ll n$. In practice, the retrieval of desired spectral
information is accomplished by constructing an orthonormal basis
$V_d \in R^{n\times d}$ of $K_d(\ket{\phi_0},A)$ and computing
eigenvalues and eigenvectors of the $d$ by $d$ projected matrix
$H_d = {P_{V_d}}^TAP_{V_d}$, where $P_{V_d}$ is projection
operator to $d$-dimensional subspace spanned by the basis $V_d$.

The most popular algorithm for finding an orthonormal basis for
the Krylov subspace, is Lanczos algorithm. The Lanczos algorithm
transforms a Hermitian matrix $A$ into a tridiagonal form
iteratively, i.e., the matrix $A$ will be of tridiagonal form in
the $d$-dimensional projected subspace $H_d$. In fact, the Lanczos
algorithm is deeply rooted in the theory of orthogonal
polynomials, which builds an orthonormal sequence of vectors
$\{\ket{\phi_0},\ket{\phi_1},...,\ket{\phi_{d-1}}\}$ and satisfy
the following three-term recursion relations
\begin{equation}\label{trt}
A\ket{\phi_i}=\beta_{i+1}\ket{\phi_{i+1}}+\alpha_i\ket{\phi_i}+\beta_i\ket{\phi_{i-1}}.
\end{equation}

The vectors $\ket{\phi_i}, i=0,1,...,d-1$ form an orthonormal basis
for the Krylov subspace $K_d(\ket{\phi_0},A)$. In these basis, the
matrix $A$ is projected to the following symmetric tridiagonal
matrix: \begin{equation}\label{adj}A\doteq\left(
\begin{array}{cccccc}
 \alpha_0 & \beta_1 & 0 & ... &... \\
      \beta_1 & \alpha_1 & \beta_2 & 0 &... \\
      0 & \beta_2 & \alpha_2 & \beta_3 & ... \\
      \vdots& \vdots & \ddots & \ddots & \ddots\\
     ... & 0 &\;\;\ \beta_{_{d-2}}& \;\;\ \alpha_{_{d-2}} &\;\ \beta_{d-1} \\
     ...& ... &0 & \beta_{d-1} & \alpha_{d-1}\\
\end{array}
\right),\end{equation}  where the scalars $\beta_{i+1}$ and
$\alpha_i$ are computed to satisfy two requirements, namely that
$\ket{\phi_{i+1}}$ be orthogonal to $\ket{\phi_i}$ and that
$\|\ket{\phi_{i+1}}\|= 1$.

In fact, the Lanczos algorithm is a modified version of the
classical Gram-Schmidt orthogonalization process. As it can be
seen, at its heart is an efficient three-term recursion relation
which arises because the matrix
$A$ is real and symmetric.\\
If we define the Krylov matrix $K$ such that the columns of $K$
are Krylov basis $\{A^i\phi_0 ; i=0,...,d-1\}$ as:
$$K:=(\ket{\phi_0}, A\ket{\phi_0}, ... , A^{d-1}\ket{\phi_0}),$$
the application of the orthonormalization process to the Krylov
matrix is equivalent to the construction of an upper triangular
matrix $P$ such that the resulting sequence $\Phi=KP$ satisfies
$\Phi^\dag \Phi=1$. We denote by $\ket{\phi_j}$ and $P_j$
respectively the $j$-th column of $\Phi$ and $P$. Then we have
\begin{equation}
\braket{\phi_0}{P_i^{\dagger}(A)P_j(A)|\phi_0}=\braket{KP_i}{KP_j}=\braket{\phi_i}{\phi_j},
\end{equation}
where $P_i=a_0+a_1A+...+a_iA^i$  is a polynomial of
 degree $i$ in indeterminate $A$.

 In the remaining part of this section   we give an algorithmic outline of the Lanczos
 algorithm, where it will be used in calculation of parameters
 $\alpha_i$ and $\beta_i$ appeared in (\ref{trt}).\\
\textbf{Lanczos algorithm } \\
Input: Matrix $A\in R^{n\times n}$, starting vector
$\ket{\phi_0}$,
$\|\ket{\phi_0}\|=1$, scalar $d$\\
Output: Orthogonal basis $\{\ket{\phi_0},...,\ket{\phi_{d-1}}\}$
of Krylov subspace $K_d(\ket{\phi_0},A)$
$$
\beta_0=0, \ket{\phi_0}=\ket{\phi}/\|\ket{\phi}\|
$$
 $$ for\;\;\ i=0,1,2,...$$
$$
\ket{\upsilon_i}=A\ket{\phi_i}
$$
$$
\alpha_i=\braket{ \phi_i}{\upsilon_i}
$$
$$
\ket{\upsilon_{i+1}}=\ket{\upsilon_i}-\beta_{i}\ket{\phi_{i-1}}-\alpha_i\ket{\phi_i}
$$
$$
\beta_{i+1}=\|\ket{\upsilon_{i+1}}\|
$$
$$ if$$
$$
\beta_{i+1}\neq 0
$$
$$
\ket{\phi_{i+1}}=\ket{\upsilon_{i+1}}/\beta_{i+1}
$$
$$else$$
$$\ket{\phi_{i+1}}=0.$$
\subsection{Spectral distribution method}
For any pair $(A,\ket{\phi_0})$ of a matrix $A$ and a vector
$\ket{\phi_0}$, one can assign a measure $\mu$ as follows
\begin{equation}\label{sp1}
\mu(x)=\braket{ \phi_0}{E(x)|\phi_0},
\end{equation}
 where
$E(x)=\sum_i|u_i\rangle\langle u_i|$ is the operator of projection
onto the eigenspace of $A$ corresponding to eigenvalue $x$, i.e.,
\begin{equation}
A=\int x E(x)dx
\end{equation}
so that, for any polynomial $P(A)$ we have
\begin{equation}\label{sp2}
P(A)=\int P(x)E(x)dx,
\end{equation}
where for discrete spectrum the above integrals are replaced by
 summation.
For example, in continuous time quantum walk on a graph \cite{js},
the expectation value of powers of the corresponding adjacency
matrix $A$ over starting site $\ket{\phi_0}$ can be written as
\begin{equation}\label{v2}
\braket{\phi_{0}}{A^m|\phi_0}=\int_{R}x^m\mu(dx), \;\;\;\;\
m=0,1,2,....
\end{equation}
The existence of a spectral distribution satisfying (\ref{v2}) is
a consequence of Hamburger's theorem, see e.g., Shohat and
Tamarkin [\cite{st}, Theorem 1.2].

From the orthogonality of vectors $\ket{\phi_j}$ (Krylov bases)
produced from Lanczos algorithm process we have,
$$\delta_{ij}=\braket{\phi_i}{\phi_j}=\braket{\phi_0}{P_i^{\dagger}(A)P_j(A)|\phi_0}$$
\begin{equation}
=\int P_i^*(x)P_j(x)\mu(x)dx=(P_i,P_j)_{\mu}.
\end{equation}
 Conversely if $P_0,...,P_{d-1}$ is the system of orthonormal
polynomials for the measure $\mu$ then the vectors
\begin{equation}\label{poly}
\ket{\phi_j}= P_j(A)\ket{\phi_0},
\end{equation}
will coincide with the sequence of orthonormal vectors produced by
the Lanczos algorithm applied to $(A,\ket{\phi_0})$.

Now, substituting (\ref{poly}) in (\ref{trt}), we get three term
recursion relations between polynomials $P_j(A)$, which leads to
 the following  three term recursion between polynomials $P_j(x)$
\begin{equation}\label{eq6}
\beta_{k+1}P_{k+1}(x)=(x-\alpha_k)P_{k}(x)-\beta_kP_{k-1}(x)
\end{equation}
for $k=0,...,d-1$. Multiplying by $\beta_1...\beta_k$ we obtain
\begin{equation}
\beta_1...\beta_{k+1}P_{k+1}(x)=(x-\alpha_k)\beta_1...\beta_kP_{k}(x)-\beta_k^2.\beta_1...\beta_{k-1}P_{k-1}(x).
\end{equation}
By rescaling $P_k$ as $Q_k=\beta_1...\beta_kP_k$, the three term
recursion relations (\ref{eq6}) are replaced by
$$ Q_0(x)=1, \;\;\;\;\;\
Q_1(x)=x,$$
\begin{equation}\label{op}
Q_{k+1}(x)= (x-\alpha_{k})Q_k(x)-\beta_k^2Q_{k-1}(x),
\end{equation}
for $k=1,2, ..., d$.

In the next section, we will need the distinct eigenvalues of
adjacency matrix of a given undirected graph which can be written in
the form (\ref{adj}), and the corresponding eigenvectors in order to
obtain the minimum time at which the marked state can be reached. As
it is known from spectral theory, we have the eigenvalues $x_i$ of
the adjacency matrix $A$ as roots of the last polynomial
$Q_{d+1}(x)$ in (\ref{op}), and the normalized eigenvectors as
\cite{tsc, lipkin}
\begin{equation}\label{eigvec}\ket{\psi_i}=\frac{1}{\sqrt{\sum_{l=0}^dP^2_l(x_i)}}\left(\begin{array}{c}
                                                          P_0(x_i) \\
                                                              P_1(x_i) \\
                                                              \vdots \\
                                                              P_d(x_i) \\
                                                            \end{array}\right).
                                                            \end{equation}
\section{Quantum search via local adiabatic evolution}
In this section, we investigate quantum search in an structured
database by using the time-dependent Hamiltonian approach and
spectral distribution method. To do so, first, we recall briefly
the time-dependent Hamiltonian approach and local adiabatic
evolution employed in Ref.\cite{Roland} to Grover's problem.

Consider the evolution of a quantum system state $\ket{\psi(t)}$
subject to a time-dependent Hamiltonian $H(t)$ is described by the
Schr\"{o}dinger equation ($\hbar=1$)
\begin{equation}
i\frac{d}{dt}\ket{\psi(t)}=H(t)\ket{\psi(t)}.
\end{equation}
According to the adiabatic theorem \cite{adia}, the state of the
system will stay close to the instantaneous ground state of the
Hamiltonian at each time $t$, if the Hamiltonian varies slowly
enough. In other words, for the instantaneous energy eigenbasis
defined by $H(t)\ket{E_n(t)} = E_n(t)\ket{E_n(t)}$, if we define
the minimum gap between the lowest two eigenvalues as
\begin{equation}\label{eq1}
g_{min}=\min_{0\leq t\leq T}[E_1(t)-E_0(t)]
\end{equation}
and the maximum value of the matrix element of $\frac{dH}{dt}$
between the two corresponding eigenstates as
\begin{equation}
D_{max}=\max_{0\leq t\leq T}| \langle \frac{dH}{dt}\rangle_{1,0}|
\end{equation}
with $\langle \frac{dH}{dt}\rangle_{1,0}=\langle
E_1(t)|\frac{dH}{dt}|E_0(t)\rangle$, then the adiabatic evolution
theorem guaranties that for the system which is prepared in its
ground state $|E_0(0)\rangle$  at time $t=0$, and evolved under the
Hamiltonian $H(t)$, we will have
\begin{equation}
| \langle E_0(T)|\psi(T)\rangle|^2\geq 1-\epsilon^2
\end{equation}
provided that
\begin{equation}\label{eq2}
\frac{D_{max}}{g^2_{min}}\leq \epsilon,
\end{equation}
where $\epsilon\ll 1$.
\subsection{Quantum search in an unsorted database with local adiabatic evolution}
Assuming a set of $N$ items in an unsorted database (uniform
distribution) where one of them is marked, the main goal is finding
the marked item in a minimum run time. To this end, a quantum state
$\ket{i}$ is assigned to each item $i$, so that the state space of
the database is spanned by the states $\ket{i}$ with $i=0,1,\ldots,
N-1$, where the marked state is denoted by $\ket{m}$. Since
$\ket{m}$ is unknown a priori, the initial state is chosen as an
equal superposition of all basis states, i.e.,
\begin{equation}
\ket{\psi_0}=\frac{1}{\sqrt{N}}\sum_{i=0}^{N-1}\ket{i}.
\end{equation}
Then, the Hamiltonian of the system is initially chosen as
\begin{equation}
H_0=I_N-\ket{\psi_0}\langle \psi_0|
\end{equation}
whose ground state is $\ket{\psi_0}$ with energy zero. Since the
database is unsorted, we can associate the items of the database
with the nodes of the complete graph with $N$ vertices (denoted by
$K_N$), for which the adjacency matrix is given by $A=J_N-I_N$.
Therefore the initial Hamiltonian can be written as
\begin{equation}\label{eq7}
H_0=I_N-\frac{1}{N}\sum_{i,j=0}^{N-1}\ket{i}\langle
j|=I_N-\frac{1}{N} J_N=\frac{1}{N}[(N-1)I_N-A]=\frac{1}{N}L,
\end{equation}
where, $J_N$ is an $N\times N$ all one matrix and $L=(N-1)I_N-A$
is the Laplacian of the complete graph $K_N$. Now, suppose that we
are able to apply to our system the Hamiltonian
\begin{equation}
H_m=I_N-\ket{m}\langle m|,
\end{equation}
whose ground state is the unknown marked state $\ket{m}$. Then, the
corresponding time dependent Hamiltonian is considered as the
following linear interpolation between $H_0$ and $H_m$
\begin{equation}\label{eq7'}
H(s)=(1-s)H_0+sH_m=\frac{1}{N}[(N-1+s)I_N-(1-s)A-Ns\ket{m}\langle
m|],
\end{equation}
where $s$ is an evolution function of time $t$ and must be found
optimally with the boundary conditions $s(0)=0$ and $s(T)=1$. In
the global adiabatic evolution $s(t)$ is chosen as a linear
function of $t$ as $s(t)=t/T$, but in the local adiabatic
evolution the time interval $T$ is divided into infinitesimal time
intervals $dt$ and the adiabaticity condition is locally applied
to each of these intervals. Applying Eq.(\ref{eq2}) to each
infinitesimal time intervals, the local adiabatic condition is
given by
\begin{equation}\label{eq3'}
|\frac{ds}{dt}|\leq \epsilon \frac{g^2(s)}{| \langle
\frac{dH}{dt}\rangle_{1,0}|}
\end{equation}
for all times $t$. Now, by using the fact that $| \langle
\frac{dH}{dt}\rangle_{1,0}|\leq 1$, the condition (\ref{eq3'}) is
verified provided that
\begin{equation}\label{eq3''}
|\frac{ds}{dt}|=\epsilon g^2(s).
\end{equation}

The algorithm consists in preparing the system in the state
$\ket{\psi_0}$ and then applying the Hamiltonian $H(s)$ during a
time $T$  so that $s(T)=1$. In order to obtain the eigenvalues of
$H(s)$ and evaluate (\ref{eq1}), we use the Lanczos algorithm to
obtain the orthonormal basis in which the hamiltonian $H(s)$ can
be reduced to as a tridiagonal matrix. To do so, we choose
$\ket{m}\equiv \ket{\phi_0}$ as the reference state (starting
vector in the Lanczos iteration algorithm) and apply $H(s)$ on it
to obtain the vector orthogonal to $\ket{\phi_0}$ as
$$\ket{\phi_1}=\frac{1}{\sqrt{N-1}}(\sqrt{N}\ket{\psi_0}-\ket{\phi_0}).$$
Now, the Hamiltonian $H(s)$ can be represented in the new
orthonormal basis states $\ket{\phi_0}$ and $\ket{\phi_1}$ as
follows
\begin{equation}
H(s)=\frac{1}{N}\left(\begin{array}{cc}
      (1-s)(N-1) & -\sqrt{N-1}(1-s) \\
       -\sqrt{N-1}(1-s) & 1+s(N-1) \\
     \end{array}\right)
\end{equation}
The eigenvalues of $H(s)$ are given by $E_{\pm}=\frac{1}{2N}\{
N\pm \sqrt{N^2(1-2s)^2+4Ns(1-s)}\}$, so that the difference
between these two eigenvalues gives the gap $g(s)$ as
\begin{equation}
g(s)=\sqrt{\frac{N-4s(1-s)(N-1)}{N}}.
\end{equation}
We see that the minimum gap $g_{min}=\frac{1}{\sqrt{N}}$ is attained
for $s=1/2$. Now, by using the local adiabatic condition
(\ref{eq3''}) we obtain
\begin{equation}
|\frac{ds}{dt}|=\frac{\epsilon}{N} [N-4s(1-s)(N-1)].
\end{equation}
After integration, one can find
\begin{equation}
t=\frac{1}{2\epsilon}\frac{N}{\sqrt{N-1}}\{\arctan[\sqrt{N-1}(2s-1)]+\arctan{\sqrt{N-1}}\}.
\end{equation}
One may now evaluate the computation time of the algorithm by taking
$s = 1$. For $N \gg 1$, we obtain
\begin{equation}
T\simeq\frac{\pi}{2\epsilon}\sqrt{N},
\end{equation}
which gives a quadratic speed-up with respect to a classical
search, so that the algorithm can be viewed as the adiabatic
evolution version of Grover's algorithm.
\subsection{Quantum search in a structured (non-uniform) database}
In the case that our states distributed non-uniformly, the state
space of the database can be considered as an arbitrary connected
graph other than the complete graph. Therefore, a straightforward
generalization of the relations (\ref{eq7})-(\ref{eq7'}) leads us to
consider
\begin{equation}
H_0=\gamma L,\;\;\ H_m=I-\ket{m}\langle m|
\end{equation}
so that
\begin{equation}\label{eq4}
\tilde{H}(s)=\gamma(1-s)L+s(I-\ket{m}\langle m|)=a
I_N-\gamma(1-s)A-s\ket{m}\langle m|,
\end{equation}
where $L$ is the Laplacian of the graph and $\gamma$ is a constant
parameter which is determined in such a way that the search time be
minimum (the search algorithm be optimal). We will consider regular
undirected graphs so that the corresponding adjacency matrices are
symmetric, and moreover the Laplacian $L$ and the adjacency matrix
$A$ of the graphs differ from each other in only multiple of
identity matrix, i.e., we have $L=DI-A$ with $D$ as the degree of
each vertex. So, the parameter $a$ in (\ref{eq4}) is given by
$a=\gamma D(1-s)+s$. Since we need the minimum gap between two
lowest eigenvalues of the hamiltonian $\tilde{H}(s)$, so the first
term in (\ref{eq4}) (multiple of identity) can be dropped.

By choosing $\ket{m}\equiv \ket{\phi_0}$ as the starting state for
the Lanczos iteration algorithm, the corresponding Krylov bases are
obtained via the three term recursion relation (\ref{trt}). That is
we have,
$$
\tilde{H}(s)|\phi_0\rangle=(\gamma
(s-1)\alpha_0-s)|\phi_0\rangle+\gamma (s-1)\beta_{1}|\phi_1\rangle,
$$
\begin{equation}
\tilde{H}(s)\ket{\phi_i}=\gamma(s-1)\{\beta_i\ket{\phi_{i-1}}+\alpha_i\ket{\phi_i}+\beta_{i+1}\ket{\phi_{i+1}}\}\;\
; \;\ i=1,2,\ldots, d.
\end{equation}
In other words, in the krylov bases $\ket{\phi_i}$, the Hamiltonian
$\tilde{H}(s)$ is rewritten as
\begin{equation}
\tilde{H}(s)= \gamma(s-1)H_0-sH_1
\end{equation}
where,
\begin{equation}
H_0\equiv A=\left( \begin{array}{cccccc}
               \alpha_0 & \beta_1 & 0 & 0 & \ldots & 0 \\
                      \beta_1 & \alpha_1 & \beta_2 & 0 & \ldots & 0 \\
                      0 & \beta_2 & \ddots & \ddots & \ddots & \vdots \\
                      0 & 0 & \ddots& \ddots & \ddots & 0 \\
                      \vdots & \vdots & \ddots & \ddots & \ddots & \beta_{d-1}\\
                      0 & 0 & \ldots & 0 & \beta_{d-1} & \alpha_{d-1}\\
                    \end{array}\right), \;\
H_1=\left(\begin{array}{ccccc}
1 & 0 & \ldots & \ldots & 0 \\
0 & 0 & 0 & \ldots & 0 \\
\vdots & 0 & 0 & \ddots & 0 \\
0 & \vdots & \ddots & \ddots & 0 \\
0 & 0 & \ldots & 0 & 0 \\
\end{array}\right).
\end{equation}

Now, using (\ref{eigvec}) and the form of $H_1$, one can easily
evaluate
$$\langle \psi_j|H_1|\psi_i\rangle=\frac{1}{\sqrt{\sum_{l,l'=0}^dP^2_l(x_j)P^2_{l'}(x_i)}},$$
so that we obtain
$$\hspace{-6.5cm}E_i^{(1)}=\langle \psi_i|H_1|\psi_i\rangle=\frac{1}{\sum_{l=0}^dP^2_l(x_i)},$$
$$E_i^{(2)}=\sum_{j;j\neq i}\frac{|\langle \psi_j|H_1|\psi_i\rangle|^2}{E_i^{(0)}-E_j^{(0)}}=\frac{1}{\gamma(s-1)}\sum_{j;j\neq i}\frac{1}{(x_i-x_j)\sum_{l,l'=0}^dP^2_l(x_i)P^2_{l'}(x_j)}.$$
Then, the approximate eigenvalues of $\tilde{H}(s)$ up to second
order are given by
$$E_i\cong \gamma(s-1)x_i-sE_i^{(1)}+s^2E_i^{(2)},$$
such that, the energy gap $g(s)$ is given by
$$g(s)=E_1-E_0\cong
\gamma(s-1)(x_1-x_0)-s(\frac{1}{\sum_{l=0}^{d}P_l^2(x_1)}-\frac{1}{\sum_{l=0}^{d}P_l^2(x_0)})+$$
\begin{equation}\label{eq50}\frac{s^2}{\gamma(s-1)}\{\sum_{j\neq1}\frac{1}{(x_1-x_j)\sum_{l,l'=0}^{d}P_l^2(x_1)P_l'^2(x_j)}-\sum_{j\neq0}\frac{1}{(x_0-x_j)\sum_{l,l'=0}^{d}P_l^2(x_0)P_l'^2(x_j)}\}.
\end{equation}
Denoting the terms in the parentheses of the second term in
(\ref{eq50}) by $A$ and the terms in the bracket of the last term by
$B$, the energy gap is written as:
\begin{equation}\label{eq5'}g(s)\cong \gamma(s-1)(x_1-x_0)-sA +\frac{s^2}{\gamma(s-1)}B.\end{equation}
Now, in order to obtain the critical value of $\gamma$ in which the
gap $g(s)$ is minimum, we take the derivative of $g(s)$ with respect
to $\gamma$ so that $\frac{\partial g(s)}{\partial \gamma}=0$. Then,
one can obtain
\begin{equation}\label{gama}\gamma_{_{crit.}}=\frac{s}{1-s}\sqrt{\frac{B}{(x_1-x_0)}},\;\;\ s\neq
0,1.\end{equation}

By substituting $\gamma_{_{crit.}}$ in (\ref{eq5'}), we obtain
\begin{equation}\label{eq5''}g_{_{min}}(s)\cong - s[A+ 2\sqrt{B(x_1-x_0)}].\end{equation}
Now, by using (\ref{eq3''}), the search time is evaluated as
$$t \cong \int \frac{ds}{\epsilon s^2(A+ 2\sqrt{B(x_1-x_0)})^2}=\frac{1}{\epsilon s(A+
2\sqrt{B(x_1-x_0)})^2}.$$ By substituting $s=1$, the total search
time at which the marked state $|m\rangle$ is found, is given by
\begin{equation}\label{time}
T\cong \frac{1}{\epsilon (A+2\sqrt{B(x_1-x_0)})^2}.
\end{equation}
As the above result indicates, in order to obtain the search time,
one needs to evaluate the terms $A$ and $B$ defined below of
(\ref{eq50}) in terms of the corresponding graph polynomials
$P_i(x)$. On the other hand, $A$ and $B$ can be calculated easily
via a systematic computer program for calculating the corresponding
polynomials from the three term recursion relations (\ref{eq6}).
\section{Examples}
\textbf{1. Dihedral group graph}\\
 The dihedral group $G=D_{2n}$ is generated by two generators $a$ and $b$ as follows:
$$D_{2n}=<a,b:a^n=1,b^2=1,b^{-1}ab=a^{-1}>.$$
We consider the even $n=2m$, where the odd $n$ can be considered
similarly. In this case, the conjugacy classes of the group, are
given by $$C_0=\{e\} , C_i=\{a^i,a^{-i}\}; i=1,...,m-1 , C_m=\{a^m\}
, C_{m+1}=\{a^{2j}b; 0\leq j\leq m-1\} $$
$$ C_{m+2}=\{a^{2j+1}b; 0\leq j\leq m-1\},$$ so we have $m+3$ conjugacy
classes. It is well known that the adjacency matrices defined as
$A_i=\sum_{g\in C_i}R(g)$, are correspond to the underlying graph of
the so-called group association scheme \cite{scheme, pst} associated
with the group $D_{2n}$, where $R(g)$ is the regular representation
of the element $g$ of the group \cite{gordon}. Therefore, we have
$$A_0=I_2\otimes I_n ,\;\ A_i=I_2\otimes (S^i + S^{-i}) , i=1,2,...,m-1, $$
$$A_m=I_2\otimes S^m ,\;\  A_{m+1}=\sigma_x \otimes(I_n+S^2+...+S^{(m-1)}),$$
$$A_{m+2}=\sigma_x \otimes (S+S^3+...+S^{2(m-1)})$$
where, $S$ is the shift or circulant matrix defined as
$$S=\left(\begin{array}{ccccc}
    0 & 1 & 0 & \ldots   & 0 \\
    0 & 0 & 1 & \ddots & 0 \\
    0 & 0 & 0 & \ddots & \vdots \\
   \vdots & \ddots & \ddots & \ddots & 1 \\
    1 & 0 & 0 & \cdots & 0
  \end{array}\right),$$
  and $\sigma_x$ is the Pauli matrix. In order that we obtain a connected undirected graph, the adjacency matrices must be symmetrized, i.e., $A_i=A_i^t$. To this end,
  we introduce the new adjacency matrix $A'$ as $ A'=A_{m+1}+A_{m+2}=\sigma_x \otimes (I+S+...+S^{n-1})$ so that the
  Krylov bases are given by
 $$\hspace{-4cm} \ket{\phi_0}=\{\ket{e }\},$$
 $$ \hspace{-.5cm}\ket{\phi_1}= \frac{1}{\sqrt n}\{\ket{b }+\ket{ba }+\ket{ba^{n-1}}\},$$
 $$ \ket{\phi_2}= \frac{1}{\sqrt{n-1}}\{\ket{a }+\ket{a^2 }+\ket{a^{n-1}}\}$$
 and the adjacency matrix is represented as
 $$ A'=\left(
     \begin{array}{ccc}
       0 &\sqrt n & 0 \\
       \sqrt n & 0 & \sqrt {n(n-1)} \\
       0 & \sqrt {n(n-1)} & 0 \\
     \end{array}
   \right).$$
   By replacing $\alpha_0=\alpha_1=\alpha_2=0 ; \beta_1=\sqrt{n},
   \beta_2=\sqrt{n(n-1)}$ in the three term recursion relation (\ref{op}),
   the corresponding polynomials are obtained as
   $$Q_0(x)=1, \;\;\ Q_1(x)=x, \;\ Q_2(x)=x^2-n,$$
   $$Q_3(x)=x(x^2-n^2).$$
   So, the corresponding eigenvalues are given by $x_0=-n$, $x_1=0$
   and $x_2=n$ (the roots of $Q_3(x)$).
Then by replacing in (\ref{eq50}), (\ref{eq5''}) and (\ref{time})
respectively, and considering $n\gg 1$, we obtain
$$A=\frac{2n-3}{2n}\cong 1, \;\;\ B=\frac{4n-3}{4n^3}\cong \frac{1}{2n^2},$$
$$\gamma_{crit.}\cong \frac{s}{1-s} \frac{\sqrt{4n-3}}{2n^{2}}, \;\;\ g_{min}(s)\cong s( \frac {2n+2\sqrt{4n-3}-3}{2n}),$$
$$ T\cong\frac{4n^{2}}{\varepsilon (2n+4\sqrt n)^{2}}.$$
  \textbf{2. m-partite Graph}\\
An $m$-partite graph (i.e., a set of graph vertices decomposed into
$m$ disjoint sets such that no two graph vertices within the same
set are adjacent) such that every pair of graph vertices in the $m$
sets are adjacent. Considering an $m$-partite graph which has $n$
vertices in each of its disjoint sets, the corresponding adjacency
matrix is given by
$$ A= K_{m}\otimes J_{n},$$
where, $ K_{m} $ is the adjacency matrix of the complete graph with
$m$ vertices and $J_{n}$ is $ n \times n $ all one matrix. Now by
using the Lanczos iteration algorithm, the Krylov bases are obtained
as
$$\hspace{-8.5cm} \ket{\phi_0}=\ket{1 },$$
$$ \ket{\phi_1}= \frac{1}{\sqrt {n(m-1)}}(\ket{n+1 }+\ket{n+2 } +\cdots +\ket{n(m-1)}),$$
$$ \hspace{-3.5cm}\ket{\phi_2}= \frac{1}{\sqrt {(n-1)}}(\ket{2 }+\ket{3 }+ \cdots +\ket{n})$$
where, the adjacency matrix is reduced to the following $3\times 3$
tridiagonal form
$$ A=\left(
     \begin{array}{ccc}
       0 &\sqrt {n(m-1)} & 0 \\
       \sqrt{ n(m-1)} & n(m-2) & \sqrt {n(n-1)(m-1)} \\
       0 & \sqrt { n(n-1)(m-1)} & 0 \\
     \end{array}
   \right).$$
Now, as illustrated in the previous example in detail, by replacing
$\alpha_0=\alpha_2=0, \alpha_1=n(m-2)$ and $\beta_1=\sqrt{n(m-1)},
\beta_2=\sqrt{n(m-1)(n-1)}$ in three term recursion relations
(\ref{op}), and evaluating the polynomials $P_i(x)$ for $i=0,1,2$
and $Q_3(x)$, one can obtain
$$x_0=-n,\;\ x_1=0,\;\ x_2=n(m-1)$$ and
$$A=\frac{m(n-1)-1}{mn},\;\;\; B=\frac{1-n}{n}[\frac{1-(m-1)^2}{n^2m(m-1)}]+\frac{1}{nm}[\frac{m^2n-2m+1}{n^2m^2(m-1)}],$$
where for $n\gg1, m\gg1$, $A$ and $B$ are approximately as
$$A\cong 1, \;\;\ B\cong \frac{1}{n^2}.$$
Then, by using (\ref{gama}), (\ref{eq5''}) and (\ref{time}), one can
obtain
$$ \gamma_{crit.}\cong\frac{s}{1-s} \sqrt \frac {1}{n^{3}(m-1)},$$
$$g_{min}(s) \cong s(1+2\sqrt \frac {m-1}{n}),$$
and
$$ T\cong\frac{1}{\varepsilon( 1+2\sqrt \frac {m-1}{n})^{2}},$$
respectively.\\\\
\textbf{3. Crown Graph}\\
A crown graph on $2n$ vertices is an undirected graph with two sets
of vertices $u_i$ and $v_i$ and with an edge from $u_i$ to $v_j$
whenever
 $i\neq j$. The adjacency matrix of this graph is given by:
$$ A= K_{n}\otimes \sigma_{x}$$
where, $ K_{n} $ is the adjacency matrix of the complete graph with
$n$ vertices and $ \sigma_{x}  $ is the Pauli matrix.  Then, the
Krylov bases are given by
$$\hspace{-8.25cm} \ket{\phi_0}=\ket{1 },$$
$$ \hspace{-1cm}\ket{\phi_1}= \frac{1}{\sqrt {n-1}}(\ket{n+1 }+\ket{n+2 } +\cdots +\ket{2n-1}),$$
$$ \hspace{-3.5cm}\ket{\phi_2}= \frac{1}{\sqrt {n-1}}(\ket{2 }+\ket{3 }+ \cdots +\ket{n}),$$
$$\hspace{-8cm} \ket{\phi_3}=\ket{2n }.$$
In the above bases, the adjacency matrix is represented as
 $$ A=\left(
        \begin{array}{cccc}
          0 & \sqrt {n-1} & 0 & 0 \\
          \sqrt {n-1} & 0 & n-2 & 0 \\
          0 & n-2 & 0 & \sqrt {n-1} \\
          0 & 0 & \sqrt {n-1} & 0 \\
        \end{array}
      \right).
 $$
Again, by using (\ref{op}) and (\ref{eq50}) one can easily calculate
the roots $x_i$ and the quantities $A$ and $B$ as follows:
$$x_0=-(n-1),\;\ x_1=-1,\;\ x_2=1, \;\ x_3=(n-1),$$
$$A\cong\frac{n-2}{2n},\;\;\; B\cong\frac{(n-1)^2}{n^2}.$$
Now, by using (\ref{gama}), (\ref{eq5''}) and (\ref{time}), one can
obtain
$$ \gamma_{crit.}\cong\frac{s}{1-s} \sqrt \frac {n-1}{n^{2}},$$
$$g_{min}(s) \cong s(1+2\sqrt \frac {(n-2)(n-1)^{2}}{n^{2}}),$$
$$ T\cong\frac{1}{\varepsilon (1+2\sqrt{ \frac {(n-2)(n-1)^{2}}{n^{2}}})^{2}}\cong\frac{1}{\varepsilon n}.$$
\section{Conclusion}
Based on the local adiabatic evolution of the Hamiltonian and some
techniques such as Krylov subspace projection methods, Lanczos
iteration algorithm and spectral distribution methods, a new
procedure for investigating quantum search in the state space of a
graph was introduced, where an approximate analytical formula for
calculating the corresponding minimum energy gap and the total
search time was given.

\end{document}